\documentclass[journal,10pt]{IEEEtran}
\usepackage{graphicx}
\usepackage{subfigure}
\usepackage{float}
\usepackage{subfloat}
\usepackage{amsmath}
\usepackage{amssymb}
\usepackage{amsthm}
\usepackage{epstopdf}
\usepackage{cases}
\usepackage{color}
\usepackage{makecell}
\usepackage{cite}
\usepackage{algorithmic}
\usepackage[ruled]{algorithm2e}
\usepackage{caption}
\usepackage{mathtools}

\usepackage[normalem]{ulem}

\makeatletter
\renewcommand\normalsize{%
	\@setfontsize\normalsize\@xpt\@xiipt
	\abovedisplayskip 6\p@ \@plus2\p@ \@minus5\p@
	\abovedisplayshortskip \z@ \@plus3\p@
	\belowdisplayshortskip 6\p@ \@plus3\p@ \@minus3\p@
	\belowdisplayskip \abovedisplayskip
	\let\@listi\@listI}
\makeatother
\ifCLASSINFOpdf
\else
\fi
\begin{document}

%
\title{Concurrent Downlink and Uplink Joint Communication and Sensing for 6G Networks}
%
%
%

\author{Xu Chen,~\IEEEmembership{Student Member,~IEEE,}
	Zhiyong Feng,~\IEEEmembership{Senior Member,~IEEE,}
	Zhiqing Wei,~\IEEEmembership{Member,~IEEE,} \\
	J. Andrew Zhang,~\IEEEmembership{Senior Member,~IEEE,}
	Xin Yuan,~\IEEEmembership{Member,~IEEE,}
	and Ping Zhang,~\IEEEmembership{Fellow,~IEEE}

\thanks{This work is supported by the National Key Research and Development Program of	China under Grants \{2020YFA0711300, 2020YFA0711302, and 2020YFA0711303\}, the National Natural Science Foundation of China under Grants \{61941102, 61790553\}, and BUPT Excellent Ph.D. Students Foundation under grant CX2021110.}
	\thanks{X. Chen, Z. Feng, Z. Wei, and Ping Zhang are with Beijing University of Posts and Telecommunications, Key Laboratory of Universal Wireless Communications, Ministry of Education, Beijing 100876, P. R. China (Email:\{chenxu96330, fengzy, weizhiqing, pzhang\}@bupt.edu.cn).}
	\thanks{J. Andrew Zhang is with the Global Big Data Technologies Centre, University of Technology Sydney, Sydney, NSW 2007, Australia	(e-mail: andrew.zhang@uts.edu.au).}
	\thanks{X. Yuan is with Commonwealth Scientific and Industrial Research Organization (CSIRO), Australia (email: Xin.Yuan@data61.csiro.au).}
	\thanks{Corresponding author: Zhiyong Feng, Zhiqing Wei}
	}

%
%

\markboth{}%
{Shell \MakeLowercase{\textit{et al.}}: Bare Demo of IEEEtran.cls for IEEE Journals}
%



\maketitle

\pagestyle{empty}  
\thispagestyle{empty} 

\newcounter{mytempeqncnt}
\setcounter{mytempeqncnt}{\value{equation}}
\begin{abstract}
	Joint communication and sensing (JCAS) is a promising technology for 6th Generation (6G) mobile networks, such as intelligent vehicular networks, intelligent manufacturing, and so on. Equipped with two spatially separated antenna arrays, the base station (BS) can perform downlink active JCAS in a mono-static setup. This paper proposes a Concurrent Downlink and Uplink (CDU) JCAS system where the BS can use the echo of transmitted dedicated signals for sensing in the uplink timeslot, while performing reliable uplink communication. A novel successive interference cancellation-based CDU JCAS processing method is proposed to enable the estimation of uplink communication symbols and downlink sensing parameters. Extensive simulation results verify the feasibility of the CDU JCAS system, showing a performance improvement of more than 10 dB compared to traditional JCAS methods while maintaining reliable uplink communication.
\end{abstract}


\section{Introduction}
Critical machine-type applications of 6th generation (6G) networks, such as intelligent vehicular networks, intelligent manufacturing and smart cities, require terminals and infrastructures to have environmental sensing and machine cooperation capabilities to facilitate automatic control of intelligent machines~\cite{Feng2021JCSC}. Therefore, wireless communication and sensing abilities are indispensable for 6G networks. However, the proliferation of wireless sensing and communication infrastructure and devices will result in severe spectrum congestion problems~\cite{liu2020joint}. In this context, joint communication and sensing (JCAS) has emerged as one of the most promising 6G key techniques due to its potential in improving spectrum and energy efficiency. It can achieve wireless sensing and communication abilities simultaneously by using the unified spectrum and transceiver with the same signal transmission~\cite{Chen2021CDOFDM}. 

Recently, the full-duplex (FD) array and transceiver designs required to implement downlink (DL) active JCAS have been widely studied. In~\cite{FDJCS2021}, the authors pointed out that the critical enabler for implementing DL active JCAS is the FD operation to simultaneously transmit JCAS signals and receive reflections from the environment. In~\cite{Andrew2021PMN}, the authors proposed that a feasible near-term solution for the FD JCAS operation is using two sets of spatially well-separated antenna arrays for transmitting and receiving. In~\cite{IBFDJCR}, the authors developed an FD JCAS system that detects targets within 20 meters while maintaining an FD link with another communication node. 

In this paper, we propose a concurrent downlink and uplink (CDU) JCAS system based on OFDM signals, which can double the sensing efficiency of BS with negligible impact on uplink (UL) communications. It exploits the two-array setup in a JCAS node, and conducts DL active sensing in the UL data communication timeslot while receiving the UL communication signals. It is particularly suitable for millimeter wave (mmWave) signals, given its propagation property of dominated line-of-sight (LoS) path.
 To our best knowledge, this is the first work that investigates this setup. The main advantage, compared to sensing using the UL signals, is that sensing can be conducted in a well-controlled manner and in a mono-static setup. Therefore, the CDU JCAS is indispensable for sensing environment consecutively. 
 In order to realize CDU JCAS, suppressing the mutual interference between UL communication and DL active sensing signals at the BS is critical. We propose a novel successive interference cancellation (SIC)-based method to remove the interference of UL communication to echo sensing signal processing without reducing the reliability of UL communication significantly. This method enables both the effective estimation of UL communication symbols and DL sensing echo channels. 


\textbf{Notations}: Bold uppercase letters denote matrices (e.g., $\textbf{M}$); bold lowercase letters denote column vectors (e.g., $\textbf{v}$); scalars are denoted by normal font (e.g., $\gamma$); $\left(\cdot\right)^H$, $\left(\cdot\right)^{*}$ and $\left(\cdot\right)^T$ denote Hermitian transpose, complex conjugate and transpose, respectively; ${\left\| {\mathbf{v}}_{k}  \right\|_l}$ represents the $l$-norm of ${\mathbf{v}}_{k}$, ${\rm{Tr}}\left( {\bf{M}} \right)$ is the trace of $\bf{M}$, and $E\left( \cdot \right)$ represents the expectation of random variables.

\section{System Model} \label{sec:System_model}
In this section, we present the CDU JCAS scenario and JCAS channel models to provide fundamentals for the CDU JCAS signal processing.

\subsection{CDU JCAS Scenario}

\begin{figure}[!t]
	\centering
	\includegraphics[width=0.28\textheight]{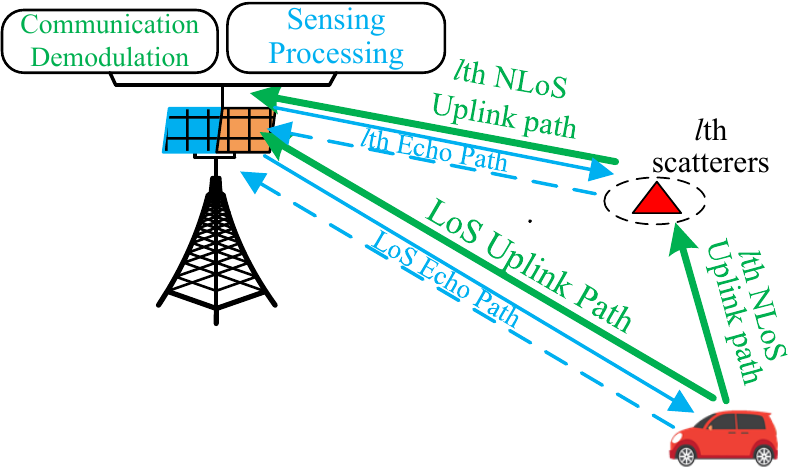}%
	\DeclareGraphicsExtensions.
	\caption{The CDU JCAS model.}
	\label{fig: Downlink JCS Model}
\end{figure}
\begin{figure}[!t]
	\centering
	\includegraphics[width=0.28\textheight]{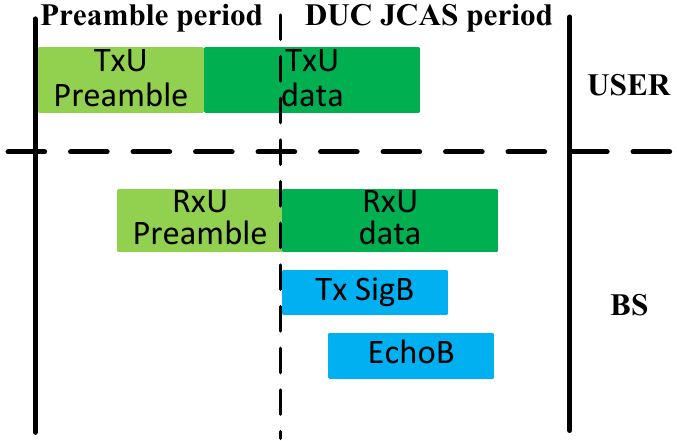}%
	\DeclareGraphicsExtensions.
	\caption{An illustration of signal structure and arrangement in the CDU JCAS scenario.}
	\label{fig:timeframe}
\end{figure}

We consider a CDU JCAS scenario, where the BS performs DL active JCAS operation while receiving the UL communication signal from the user, as illustrated in Fig.~\ref{fig: Downlink JCS Model}. Both the user and BS are equipped with uniform plane arrays (UPAs). BS has two spatially well-separated UPAs for JCAS operation, where one is used for transmitting JCAS signals, and the other concurrently receives JCAS echo signals and UL signals from the user. Our focus here is on enabling active sensing at BS in the UL communication period.

Fig.~\ref{fig:timeframe} shows the signal structure and arrangement in the CDU JCAS scenario. In the preamble period, UL communication synchronization and channel estimation are conducted at BS. Then, BS starts to transmit dedicated sensing signals, simultaneously with receiving the rest of UL communication signals. The BS soon receives superimposed UL communication and DL sensing echo signals. OFDM signals are used for both UL communication and DL active sensing. The repeated sequences are used for each DL sensing symbol so that each symbol can be treated as the cyclic prefix of the next symbol, making it easy to align the reception of UL communication and DL echo signal for OFDM signal processing. As a result, concurrent UL communication and DL active sensing are achieved by further handling mutual interference.

\subsection{UPA Model}
The uniform interval between neighboring antenna elements is denoted by $d_a$. The size of UPA is ${P} \times {Q}$. The two-dimensional (2D) angle-of-arrival (AoA) for receiving or the angle-of-departure (AoD) for transmitting the far-field signal is ${{\bf{p}}} = {( {{\varphi},{\theta}} )^T}$, where ${\varphi}$ is the azimuth angle, and ${\theta}$ is the elevation angle. The phase difference between the $(p,q)$th antenna element and the reference element is 
\begin{equation}\label{equ:phase_difference}
	{a_{p,q}} ( {{{\bf{p}}}} ) \! =\! \exp [  { - j \!\frac{{2\pi }}{\lambda }{d_a}  \!\!( {p\cos {\varphi} \sin {\theta}  \!+ \! q\sin {\varphi} \sin {\theta}} )} ],
\end{equation}
where $\lambda = c/f_c$ is the wavelength of the carrier, $f_c$ is the carrier frequency, and $c$ is the speed of light in vacuum.

The steering vector for the array is given by 
\begin{equation}\label{equ:steeringVec}
	{\bf{a}}\left( {{{\bf{p}}}} \right) = \left[ {{a_{p,q}}\left( {{{\bf{p}}}} \right)} \right]\left| {_ {(p,q) \in [ {0,1, \cdots ,P - 1}] \times [ {0,1, \cdots ,P - 1}]}}\right.,
\end{equation}
where $\mathbf{a}(\mathbf{p}) \in \mathbb{C}^{PQ \times 1}$, and ${\left. {\left[ {{v_{p,q}}} \right]} \right|_{(p,q) \in {\bf{S}}1 \times {\bf{S}}2}}$ denotes the vector stacked by values ${v_{p,q}}$ satisfying $p\in{\bf{S}}1$ and $q\in{\bf{S}}2$. 


\subsection{JCAS Channel Models}\label{sec:DUC_JCS_Channel_Model}
Without loss of generality, we assume that two arrays of BS have the same size, denoted by ${P_t} \times {Q_t}$, while the size of the user's array is ${P_r} \times {Q_r}$. In the proposed CDU JCAS scenario, the UL communication and the DL echo sensing subchannels are concerned. 

\subsubsection{Downlink Echo Sensing Subchannel Model}

The DL sensing subchannel comprises the reflected paths of the user and scatterers, as shown in Fig.~\ref{fig: Downlink JCS Model}.
The channel response of the DL echo sensing subchannel at the $m$th OFDM symbol of the $n$th subcarrier is given by
\begin{equation}\label{equ:JCS_sensing_channel}
	{\bf{H}}_{S,n,m}^D{\rm{ = }}\sum\limits_{l = 0}^{{L_T} - 1} {\left[ \begin{array}{l}
			{b_{S,l}}{e^{j2\pi {f_{s,l}}mT_s^D}}{e^{ - j2\pi n\Delta {f^D}( {{\tau _{s,l}}} )}}\\
			\times {\bf{a}}( {{\bf{p}}_{RX,l}^D} ){{\bf{a}}^T}( {{\bf{p}}_{TX,l}^D} )
		\end{array} \right]},
\end{equation}
where ${\bf{p}}_{TX,l}^D$ and ${\bf{p}}_{RX,l}^D$ are 2D AoD and AoA of the JCAS transmitter and sensing receiver, respectively; ${{\bf{a}}( {{\bf{p}}_{TX,l}^D})} \in \mathbb{C}^{{P_t}{Q_t} \times 1}$ and ${{\bf{a}}( {{\bf{p}}_{RX,l}^D})} \in \mathbb{C}^{{P_t}{Q_t} \times 1}$ are the corresponding steering vectors as given in \eqref{equ:steeringVec}; Since the mmWave array is typically small, ${\bf{p}}_{RX,l}^D = {\bf{p}}_{TX,l}^D$ holds; $T_s^D$ and $\Delta {f^D}$ are the time duration and subcarrier interval of each OFDM symbol, respectively; ${f_{s,l}} = \frac{{2{v_{l}}}}{\lambda }$ is the reflection Doppler frequency shifts of the $l$th echo path, with $v_{l}$ being the corresponding radial relative velocity; ${\tau _{s,l}} = \frac{{2{d_l}}}{c}$ is the reflection time delays of the $l$th path, with $d_l$ being the corresponding range; and ${b_{S,l}} = \sqrt {\frac{{{\lambda ^2}}}{{{{\left( {4\pi } \right)}^3}{d_{l}}^4}}} \times {\beta _{S,l}}$ with ${\beta _{S,l}}$ being the reflecting factor of the $l$th scatterer that follows a complex Gaussian distribution with zero mean and variance $\sigma _{S\beta ,l}^2$. 

\subsubsection{Uplink Communication Channel Model}
The UL communication channel comprises a LoS path between the user and BS, and multiple non-line-of-sight (NLoS) paths, as shown in Fig.~\ref{fig: Downlink JCS Model}. 
The channel response matrix of the UL communication channel at the $m$th OFDM symbol of the $n$th subcarrier is given by
\begin{equation}\label{equ:JCS communication channel}
	{\bf{H}}_{C,n,m}^U{\rm{ = }}\sum\limits_{l = 0}^{L - 1} {\left[ \begin{array}{l}
			{b_{C,l}}{e^{j2\pi \left( {{f_{c,l}}} \right)mT_s^U}}{e^{ - j2\pi n\Delta {f^U}( {{\tau _{c,l}}} )}}\\
			\times {\bf{a}}( {{\bf{p}}_{RX,l}^U} ){{\bf{a}}^T}( {{\bf{p}}_{TX,l}^U} )
		\end{array} \right]},
\end{equation}
where $l = 0$ is for the response of the LoS path and $l = 1, \cdots ,L - 1$ are for the responses of the $(L-1)$ NLoS paths; ${\bf{a}}( {{\bf{p}}_{RX,l}^U}) \in \mathbb{C}^{{P_t}{Q_t} \times 1}$ and ${\bf{a}}( {{\bf{p}}_{TX,l}^U}) \in \mathbb{C}^{{P_r}{Q_r} \times 1}$ are the receiving and transmitting steering vectors of the $l$th path, as given in~\eqref{equ:steeringVec}, respectively; ${\bf{p}}_{RX,l}^U$ and ${\bf{p}}_{TX,l}^U$ are the corresponding 2D AoA and AoD of the $l$th path for the receiver and transmitter, respectively; $T_s^U$ and $\Delta {f^U}$ are the time duration and subcarrier interval of each UL OFDM symbol, ${f_{c,0}} = \frac{{{v_{0}}}}{\lambda }$ is the Doppler frequency shift of the LoS path, ${\tau _{c,0}} = \frac{{{d_0}}}{c}$ is the time delay of the LoS path, ${{f_{c,l}}}$ and ${\tau _{c,l}}$ ($l>0$) are the aggregate Doppler frequency shift and time delay of the $l$th NLoS path, respectively; and ${b_{C,0}} = \sqrt {\frac{{{\lambda ^2}}}{{{{(4\pi {d_0})}^2}}}}$ and ${b_{C,l}} = \sqrt {\frac{{{\lambda ^2}}}{{{{( {4\pi } )}^3}{d_{l,1}}^2{d_{l,2}}^2}}}  \times {\beta _{C,l}}$ ($l>0$) are the attenuation of the LoS path and the $l$th NLoS path, respectively, with $d_{l,1}$ and $d_{l,2}$ being the ranges between the transmitter and the $l$th scatterer, and between the $l$th scatterer and the receiver, respectively. Moreover, ${\beta _{C,l}}$ is the reflection factor at the $l$th scatterer, which follows a complex Gaussian distribution with zero mean and variance $\sigma _{C\beta ,l}^2$. Due to the existence of ${b_{C,l}}$, the LoS path is much stronger than the NLoS path for mmWave.

\begin{figure}[!t]
	\centering
	\includegraphics[width=0.35\textheight]{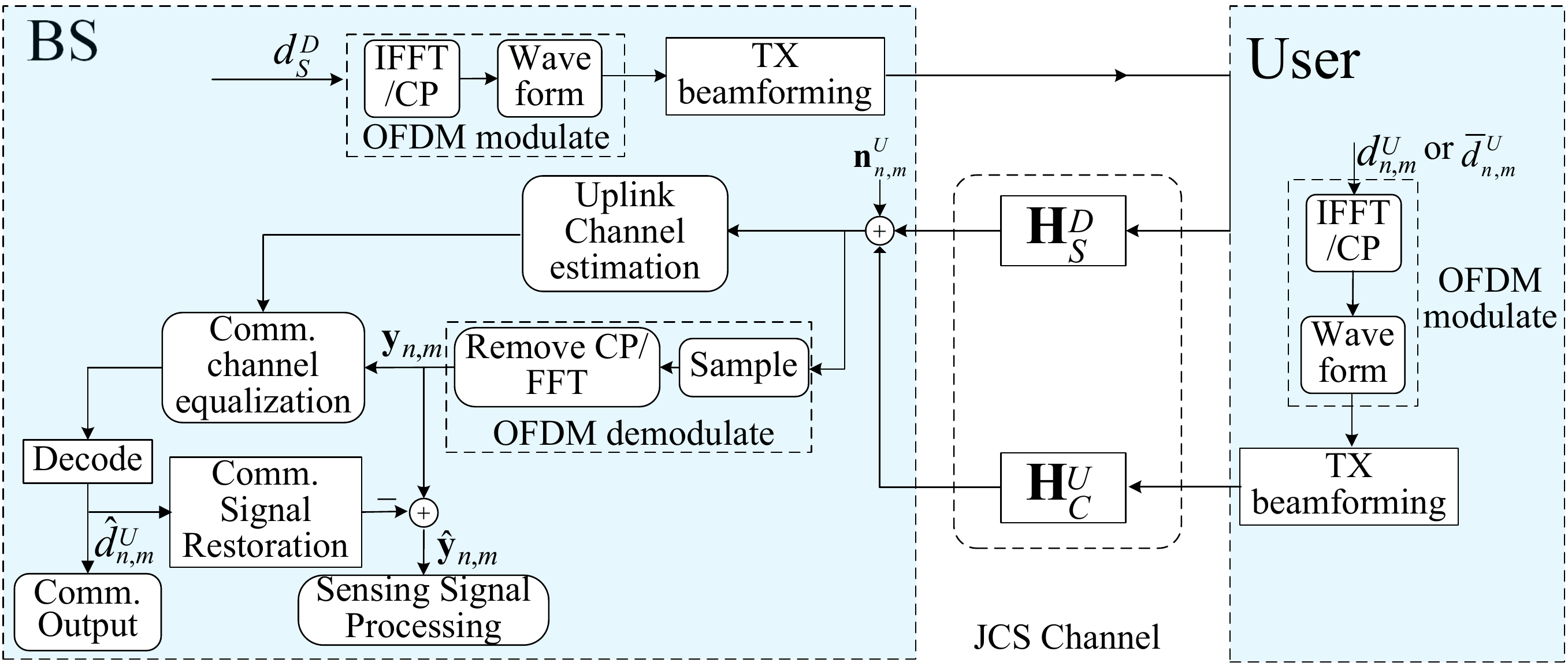}%
	\DeclareGraphicsExtensions.
	\caption{The CDU JCAS signal processing procedures.}
	\label{fig: DUC_signal_processing}
\end{figure}

\section{Concurrent Downlink and Uplink JCAS Signal Processing}\label{sec:DUC_JCS_Signal_Processing}
This section provides the CDU JCAS processing method to acquire the communication and sensing information from the superposed received signals. We consider the case where the received UL communication signal is significantly larger than the DL echo signals. This is a typical case because the path loss factor of the echo signals is larger than 4, as shown in~\eqref{equ:JCS_sensing_channel}, while it is between 2 and 3 for the UL communication signals, as shown in~\eqref{equ:JCS communication channel}. In this case, the UL communication signal is interfered by relatively smaller echo signal.  Therefore, we can demodulate the UL communication symbols first, and then reconstruct and remove the UL communication symbols from the superimposed received signals to obtain the cleaned sensing echo signals. The CDU JCAS signal processing procedures are illustrated in Fig.~\ref{fig: DUC_signal_processing}. To simplify the interference cancellation process, we let $\Delta_f^D = \Delta_f^U$ and $T_s^D = T_s^U$, which is a typical choice for DL sensing.
\subsection{UL Communication Channel Acquisition}
In the preamble period shown in Fig.~\ref{fig:timeframe}, the user transmits preamble symbols to BS. BS receives the preamble symbols expressed as
\begin{equation}\label{equ:Uplinkpreamble}
	{\bf{y}}_{n,m}^U = {\bf{H}}_{C,n,m}^U {{\bf{w}}_{TX}^U} \sqrt {P_t^U} \bar d_{n,m}^U + {\bf{n}}_{n,m}^U,
\end{equation}
where ${\bf{y}}_{n,m}^U$  is a ${P_t}{Q_t} \times 1$ vector, $\bar d_{n,m}^U$ is the preamble symbol with unit modulus, i.e., $E\{ {\| {\bar d_{n,m}^U} \|_2^2} \} = 1$, $P_t^U$ is the UL transmission power, and ${\bf{n}}_{n,m}^U$ is a Gaussian noise vector with each element following a Gaussian distribution with zero mean and variance $\sigma_N^2$. Moreover, ${\bf{w}}_{TX}^U$ is the transmit beamforming vector used to generate the beam pointing at BS. Without loss of generality, ${\bf{w}}_{TX}^U$ is generated by using the least-square (LS) beamforming method in this paper. Therefore, ${\bf{w}}_{TX}^U = {c_0}{[ {{{\bf{a}}^H}( {{\bf{p}}_{TX,l}^U} )} ]^\dag }$~\cite{Zhang2019JCRS}, where $[\textbf{A}]^\dag$ is the pseudo-inverse matrix of $\textbf{A}$, and ${c_0} = {e^{j2\pi f\phi }}$ is a complex value with unit modulus and arbitrary phase. We denote the UL transmission channel by ${\bf{h}}_{C,n,m}^U = {\bf{H}}_{C,n,m}^U{{\bf{w}}_{TX}^U} $, which is a ${P_t}{Q_t} \times 1$ matrix. The estimate of ${\bf{h}}_{C,n,m}^U$ is presented as
\begin{equation}\label{equ:UplinkCSI}
	{\bf{\hat h}}_{C,n,m}^U = {{{\bf{y}}_{n,m}^U} \mathord{/
			{\vphantom {{{\bf{y}}_{n,m}^U} {(\sqrt {P_t^U} \bar d_{n,m}^U)}}} 
			\kern-\nulldelimiterspace} {(\sqrt {P_t^U} \bar d_{n,m}^U)}}.
\end{equation}

\subsection{CDU JCAS Signal Processing}
The normalized frequency domain transmitted symbols for UL communication is denoted by $d_{n,m}^U$, which is modulated by Quadrature Amplitude Modulation (QAM)  with $E\{ \left\| {d_{n,m}^U} \right\|_2^2\}=1$. The DL active sensing symbol is denoted by $d_{n,m}^{D,S}$, which has constant modulus with $\left\| d_{n,m}^{D,S} \right\|_2^2  = 1$. Then, the received superposed signal at the $m$th OFDM symbol of the $n$th subcarrier can be given by 
\begin{equation}\label{equ:superimposed_RX}
	{\bf{y}}_{n,m}^{DU} = {\bf{h}}_{C,n,m}^U\sqrt {P_t^U} d_{n,m}^U + {\bf{h}}_{S,n,m}^D\sqrt {P_t^D} d_{n,m}^{D,S} + {\bf{n}}_{n,m}^{DU},
\end{equation}
where $P_t^D$ is the transmit power of DL sensing symbol, and ${\bf{h}}_{S,n,m}^D = {\bf{H}}_{S,n,m}^D {{\bf{w}}_{TX}^D} $ is the sensing channel response vector of dimension ${P_t}{Q_t} \times 1$. Here, ${\bf{w}}_{TX}^D$ generates the beam pointing at the detection direction of interest, denoted by ${\bf{p}}_{TX}^{DS} = ( {{\varphi_l^D},{\theta_l^D}} )$, which is also generated by LS beamforming in this paper and is expressed as ${\bf{w}}_{TX}^D = {c_0}{[ {{\bf{a}}^H( {{\bf{p}}_{TX}^{DS}} )} ]^\dag }$. Moreover, the elements in ${\bf{n}}_{n,m}^{DU}$ are independent and identically distributed, each following a complex Gaussian distribution with zero mean and variance $\sigma_N^2$. 

\subsubsection{Communication Signal Demodulation} To demodulate the communication symbol from the superimposed received signal, a receiving beamforming vector should be generated by BS, denoted by ${\bf{w}}_{RX}^C$, to equalize the UL communication channel. Here, ${\bf{w}}_{RX}^C$ is designed to maximize the signal to noise power ratio (SNR) and minimize the communication symbol demodulation error. Moreover, according to~\cite{ChenBeam2017}, the solution of ${\bf{w}}_{RX}^C$ is the linear transform of the channel CSI, i.e., ${\bf{w}}_{RX}^C = {\bf{h}}_{C,n,m}^U{\bf{B}}$. Therefore, the problem is formulated as
\begin{equation}\label{equ:argmin_pro}
		\begin{array}{*{20}{l}}
			{{\bf{w}}_{RX}^C \!=\! \arg\! \mathop {\min }\limits_{\bf{w}}\! E\!\left\{ \!\!\!\! {\begin{array}{*{20}{l}}
						{\| {{{( {\bf{w}} )}^H}{\bf{h}}_{C,n,m}^U\sqrt {P_t^U} d_{n,m}^U \!-\! d_{n,m}^U} \|_2^2}\\
						{ + {\rm{Tr}}[ {\sigma _N^2{{( {\bf{w}} )}^H}{\bf{w}}} ]}
				\end{array}} \!\!\!\! \right\}}\\
			{s.t.\;{\bf{w}} = {\bf{h}}_{C,n,m}^U{\bf{B}},}
		\end{array}
\end{equation}
Here, the received echo is ignored because ${\bf{h}}_{S,n,m}^D$ is unknown before sensing processing and is typically much smaller than communication response as aforementioned. The solution to \eqref{equ:argmin_pro} is given in (9), and the detailed derivation is provided in the \textbf{Appendix}.
\begin{equation}\label{equ:w_RX_C}
	\begin{aligned}
		{\bf{w}}_{RX}^C =& \sqrt P {\bf{h}}_{C,n,m}^U{[{( {{\bf{h}}_{C,n,m}^U} )^H}{\bf{h}}_{C,n,m}^U]^{ - 1}}\\
		&\times {[ {P{{( {{\bf{h}}_{C,n,m}^U} )}^H}{\bf{h}}_{C,n,m}^U + \sigma _N^2} ]^{ - 1}}\\
		&\times [{( {{\bf{h}}_{C,n,m}^U} )^H}{\bf{h}}_{C,n,m}^U].
	\end{aligned}
\end{equation}

After ${\bf{w}}_{RX}^C$ is acquired, the superimposed signal, $\hat r_{n,m}^{DU} = {\left( {{\bf{w}}_{RX}^C} \right)^H}{\bf{y}}_{n,m}^{DU}$, is expressed as
\begin{equation}\label{equ:r_DU}
		\hat r_{n,m}^{DU} = h_{C,n,m}^U\sqrt {P_t^U} d_{n,m}^U\!\! +\!\! h_{S,n,m}^{D,I}\sqrt {P_t^D} d_{n,m}^{D,S} \!\!+\!\! n_{n,m}^C,
\end{equation}
where $h_{C,n,m}^U = {( {{\bf{w}}_{RX}^C} )^H}{\bf{h}}_{C,n,m}^U$ and $h_{S,n,m}^{D,I} = {( {{\bf{w}}_{RX}^C} )^H}{\bf{h}}_{S,n,m}^D$ are transformed communication and sensing responses, respectively; and $n_{n,m}^C = {( {{\bf{w}}_{RX}^C} )^H}{\bf{n}}_{n,m}^{DU}$ is the transformed noise. The estimated communication response is $\hat h_{C,n,m}^U = {( {{\bf{w}}_{RX}^C} )^H}{\bf{\hat h}}_{C,n,m}^U\sqrt {P_t^U}$. Then, we can obtain the noisy communication symbol as $\bar d_{n,m}^{DU} = {{( {\hat r_{n,m}^{DU}} )} \mathord{/
		{\vphantom {{( {\hat r_{n,m}^{DU}} )} {\hat h_{C,n,m}^U}}} 
		\kern-\nulldelimiterspace} {\hat h_{C,n,m}^U}}$.
The estimate of $d_{n,m}^U$, denoted by $\hat d_{n,m}^U$, can be obtained based on $\bar d_{n,m}^{DU}$ by the maximum-likelihood (ML) criterion, which is expressed as
\begin{equation}\label{equ:d_nm_estimate}
\hat d_{n,m}^U = \mathop {\arg \min }\limits_{{d_m} \in {\Theta _{QAM}}} \| {\bar d_{n,m}^{DU} - {d_m}} \|_2^2,
\end{equation}
where ${\Theta _{QAM}}$ is the UL communication constellation set.

\subsubsection{Downlink Echo Sensing Signal Processing} 
After the communication symbol is estimated, the communication signals can be reconstructed and removed from \eqref{equ:superimposed_RX} and we obtain 
\begin{equation}\label{equ:Uplink_demodulation}
	{\bf{\hat y}}_{n,m}^{DU} = {\bf{h}}_{S,n,m}^D\sqrt {P_t^D} d_{n,m}^{D,S} + {\bf{e}}_{n,m}^U + {\bf{n}}_{n,m}^{DU},
\end{equation}
where ${\bf{e}}_{n,m}^U = {\bf{h}}_{C,n,m}^U\sqrt {P_t^U} d_{n,m}^U - {\bf{\hat h}}_{C,n,m}^U\sqrt {P_t^U} \hat d_{n,m}^U$ denotes the demodulation error, and the last two terms are interference and noise, respectively. It can be seen that noise can influence both bit error rate (BER) and communication channel estimation error.

Since we are interested in sensing at a specified direction, ${\bf{p}}_{TX}^{DS}$, BS can apply a beamforming vector pointing at this specified direction, denoted by ${{\bf{w}}_{RX,S}^D}$, to receive the cleaned echo signals in \eqref{equ:Uplink_demodulation}. Here, without loss of generality, we apply the LS beamforming method to obtain ${{\bf{w}}_{RX,S}^D}$. Because the echo AoA is the same as the sensing AoD, ${\bf{w}}_{RX,S}^D = {\left( {{\bf{w}}_{TX}^D} \right)^*}$. The echo sensing response estimated at the $n$th subcarrier of the $m$th OFDM symbol, denoted by $\hat H_{S,n,m}^D = {{{{( {{\bf{w}}_{RX,S}^D} )}^H}{\bf{\hat y}}_{n,m}^{DU}} \mathord{/
		{\vphantom {{{{( {{\bf{w}}_{RX,S}^D} )}^H}{\bf{\hat y}}_{n,m}^{DU}} {d_{n,m}^{D,S}}}} 
		\kern-\nulldelimiterspace} {d_{n,m}^{D,S}}}$, is given by~\cite{Sturm2011Waveform}
\begin{equation}\label{equ:echo_signal}
		\hat H_{S,n,m}^D \!\!=\!\!\! \sum\limits_{l = 0}^{{L_T} - 1} \!\! \left[ \!\!\begin{array}{l}
			\sqrt {P_t^D} {b_{S,l}}\varpi _{RX,l}^D\chi _{TX,l}^D\\
			\times {e^{j2\pi (m{f_{s,l}}T_s^D - n\Delta {f^D} {{\tau _{s,l}}} )}}
		\end{array} \!\!\right] \!\!+ \!\!{I_{s,n,m}} \!\!+\!\! {n_{s,n,m}},
\end{equation}
where $\chi _{TX,l}^D = {\bf{a}}^T( {{\bf{p}}_{TX,l}^D} ){( {{\bf{w}}_{TX}^D} )}$ and $\varpi _{RX,l}^D = {( {{\bf{w}}_{RX,S}^D} )^H}{\bf{a}}( {{\bf{p}}_{RX,l}^D} )$ are the transmitting and receiving beamforming gains of DL JCAS, and ${I_{s,n,m}} = {{{{({\bf{w}}_{RX,S}^D)}^H}{\bf{e}}_{n,m}^U} \mathord{/
		{\vphantom {{{{({\bf{w}}_{RX,S}^D)}^H}{\bf{e}}_{n,m}^U} {d_{n,m}^{D,S}}}} 
		\kern-\nulldelimiterspace} {d_{n,m}^{D,S}}}$ and ${n_{s,n,m}} = {{{{({\bf{w}}_{RX,S}^D)}^H}{\bf{n}}_{n,m}^{DU}} \mathord{/
		{\vphantom {{{{({\bf{w}}_{RX,S}^D)}^H}{\bf{n}}_{n,m}^{DU}} {d_{n,m}^{D,S}}}} 
		\kern-\nulldelimiterspace} {d_{n,m}^{D,S}}}$ are the transformed propagation error and noise, respectively. Note that  $\hat H_{S,n,m}^D$ is a complex scalar. 

For JCAS processing, multiple OFDM symbols and subcarriers are required. We use $M_s$ and $N_c$ to denote the numbers of OFDM symbols and subcarriers for JCAS operation, respectively. The echo sensing channel response matrix of $M_s$ echo symbols at $N_c$ subcarriers is denoted by ${\bf{\hat H}}_S^D$. The $(n,m)$th element of ${\bf{\hat H}}_S^D$ is ${[ {\bf{\hat H}}_S^D ]_{n,m}} = \hat H_{S,n,m}^D$. 

It can be seen that ${\bf{\hat H}}_S^D$ has column and row vectors with the bases ${{e^{ - j2\pi n\Delta {f^D}\left( {{\tau _{s,0}}} \right)}}}$ and ${{e^{j2\pi {f_{s,0,1}}m{T_s^D}}}}$, respectively. Therefore, by applying inverse fast Fourier transform (IFFT) to each column of ${\bf{\hat H}}_S^D$, and applying fast Fourier transform (FFT) to each row of the transformed matrix of ${{\bf{\bar H}}_S}$, the range-Doppler spectrum can be obtained according to~\cite{Sturm2011Waveform}. By denoting the coordinate of the maximal point of range-Doppler spectrum as $( {l_R,l_f} )$, the range and radial velocity are given by $\hat d_0 = \frac{{c l_R }}{{2N_c\Delta {f^D}}}$ and $\hat v_{r,0}{\rm{ = }}\frac{{\lambda l_f }}{{2M_sT_s^D}}$, respectively~\cite{Sturm2011Waveform}.

\section{Simulation Results and Analysis}\label{sec:Simulation_results_and_analysis}
In this section, we present the simulation results to verify the effectiveness of the presented CDU JCAS processing method. The simulation parameters are given as follows. The carrier frequency is 63 GHz~\cite{3GPPV2X}, the subcarrier interval is 240 kHz. The subcarrier number and OFDM symbol number for detection are $N_c = $ 128 and $M_s =$ 64, respectively. Therefore, the bandwidth of CDU JCAS is $B  $ = 30.72 MHz. The variance of complex Gaussian noise is set to $\sigma_N^2 = kFTB = 1.2294\times10^{-12} $ W, where $k$ is the Boltzmann constant, $F$ is the noise factor, and $T$ is the standard temperature.  The antenna interval is half the wavelength, and the antenna array sizes of BS and the user are $P_t \times Q_t = 8 \times 8$ and $P_r \times Q_r = 1\times 1$, respectively. The locations of BS and the user are $(50, 4.75, 7)$ m and $(140, 0, 2)$ m, respectively, and the target's location is $(129, 10, 5)$ m. We focus on range detection, and the user is assumed to be static. Moreover, we set $\sigma _{C\beta ,l}^2 = \sigma _{S\beta ,l}^2 = 1$. Based on the locations of the user, BS, and the scatterer, the AoA, AoD and ranges between the user and BS can be derived to generate the JCAS channel response matrices according to \eqref{equ:JCS_sensing_channel} and \eqref{equ:JCS communication channel}. 

The range detection mean square error (MSE) is defined as the mean value of the squared error of all the range estimates under a certain set of parameters. To show the improvement of sensing accuracy under mutual interference between UL communication and sensing echoes, we choose the DL JCAS method in \cite{Zhang2019JCRS} as a comparison, which only focuses on the sensing echo signal processing and does not consider handling UL communication interference. 

For the simplicity of demonstration, we predefine the DL JCAS method in \cite{Zhang2019JCRS} as \textit{case} 1, the proposed CDU JCAS processing method as \textit{case} 2. Moreover, we predefine the proposed CDU JCAS processing method without UL communication power as \textit{case} 3, which is a bottom line for range detection MSE.

\begin{figure}[!t]
	\centering
	\includegraphics[width=0.33\textheight]{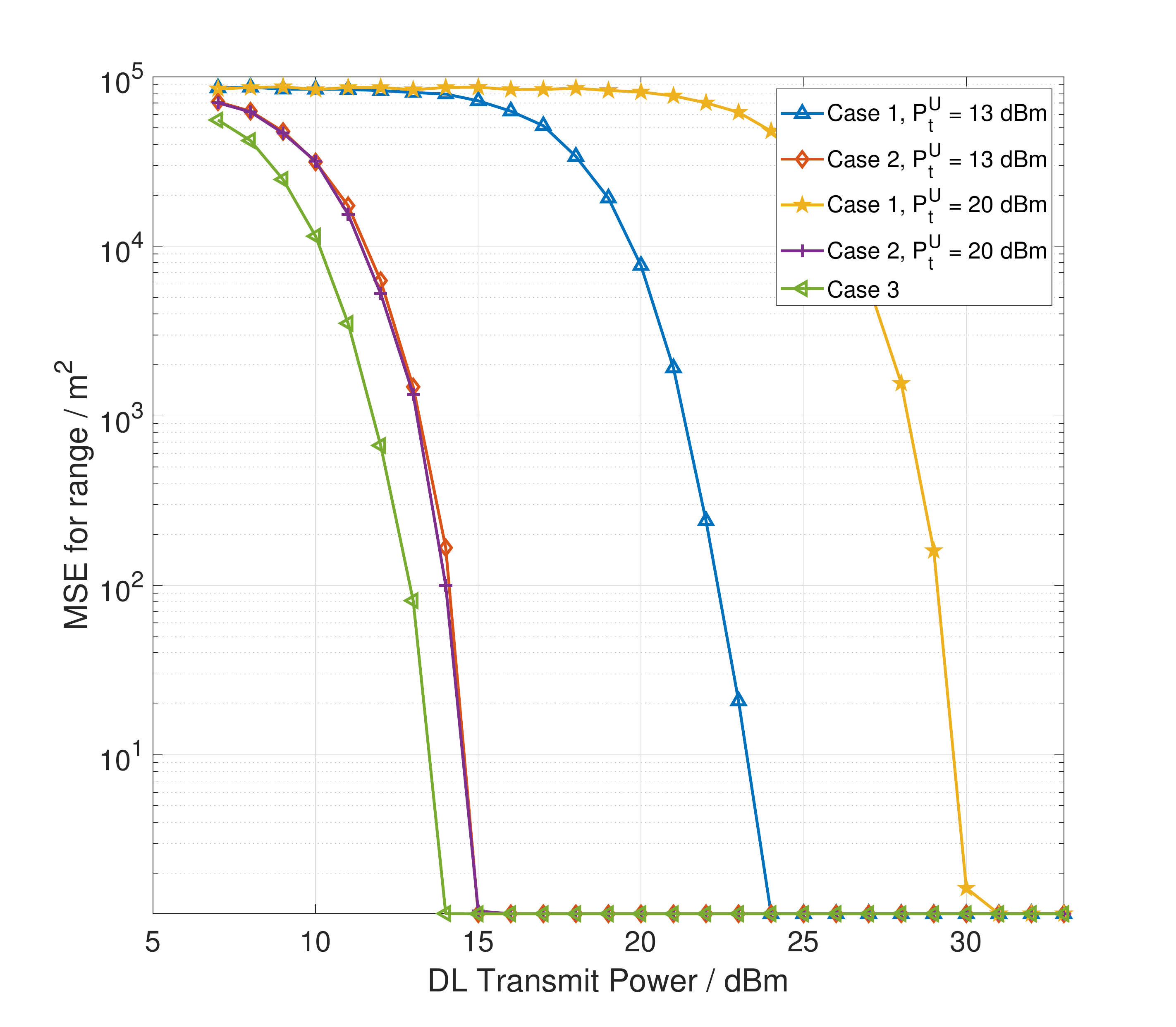}%
	\DeclareGraphicsExtensions.
	\caption{The range detection MSE of the proposed CDU JCAS method and the traditional DL JCAS method under UL communication interference.}
	\label{fig: DUC_MSE_position}
\end{figure}

\begin{figure}[!t]
	\centering
	\includegraphics[width=0.33\textheight]{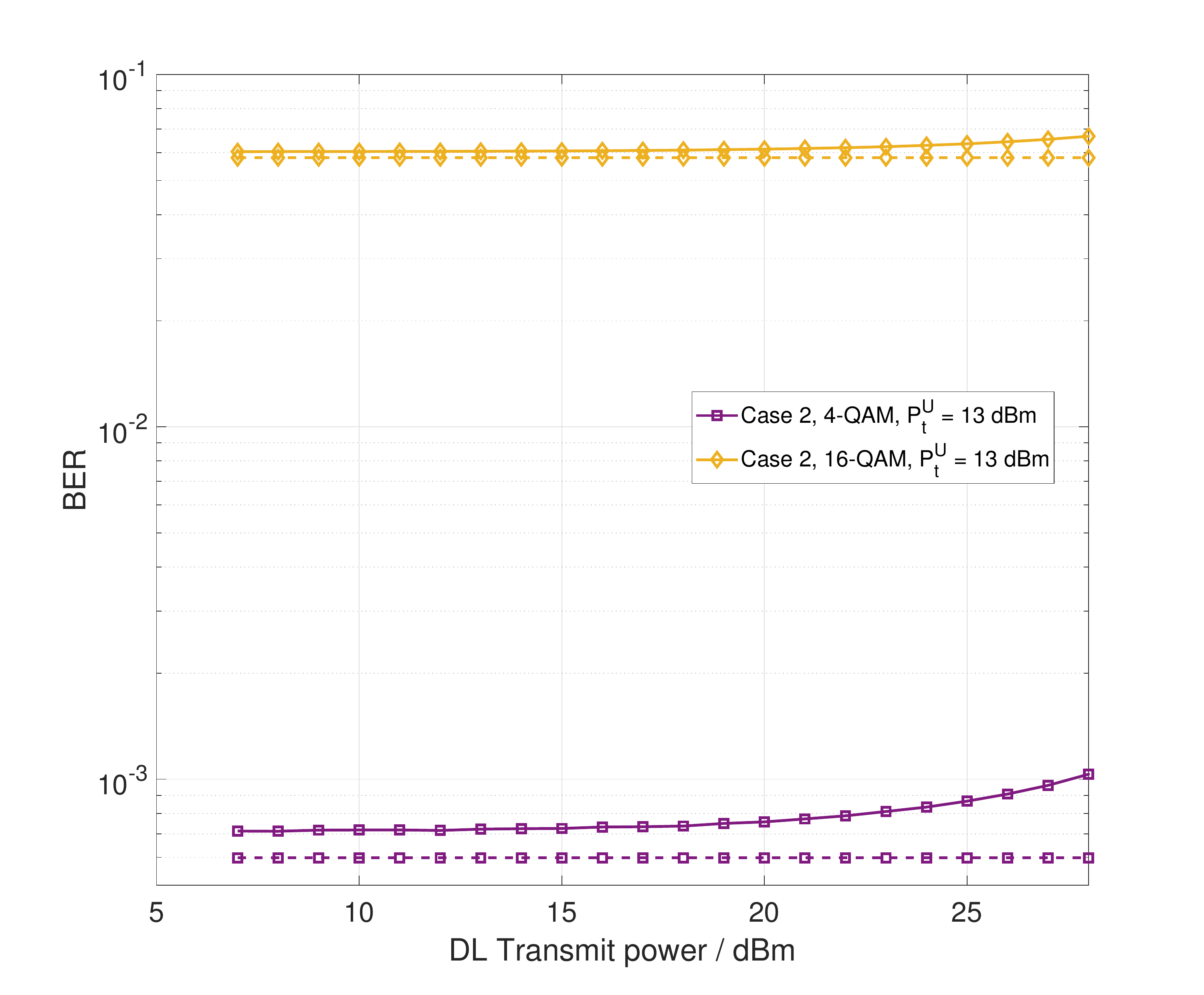}%
	\DeclareGraphicsExtensions.
	\caption{The BER of CDU JCAS processing method. The solid curves are for \textit{case} 2, and the dashed curves are for UL communication with no echo interference. Note that when $P_t^U =$ 20 dBm, the BER approaches zero and hence is not shown in the figure.}
	\label{fig: BER}
\end{figure}


Fig.~\ref{fig: DUC_MSE_position} shows the range detection MSEs of \textit{case} 1, \textit{case} 2 and \textit{case} 3 in the presence of UL communication interference when 4-QAM is used for UL communication. The UL transmit power, $P_t^U$, is set to 13 dBm and 20 dBm, respectively. As the DL transmit power of BS increases, the MSEs of all the curves decrease because the received SNRs increase. The transmit power required to achieve the minimum MSE for \textit{case} 1 is increased by more than 10 dB, compared with \textit{case} 3. In other words, the range accuracy of the traditional JCAS method is deteriorated severely in the presence of UL communication interference. 
When $P_t^U$ = 20 and 13 dBm, the transmit power required for \textit{case} 2 to achieve the minimum MSE are reduced by 16 and 9 dB, compared with \textit{case} 1. This is because CDU JCAS can reduce the impact of UL communication interference on the echo sensing processing. Moreover, when $P_t^U$ = 20 and 13 dBm, the transmit power required for \textit{case} 2 to achieve the minimum MSE are increased by 1 dB compared with \textit{case} 3. In other words, the error propagation caused by UL communication deteriorates the sensing performance.


Fig.~\ref{fig: BER} shows BER of \textit{case} 2. The BER of UL communication with no echo interference is also plotted for comparison. Note that when $P_t^U =$ 20 dBm, the BER is negligible and approaches 0, which cannot be shown in the figure. When the DL transmit power, $P_t^D$, is small, BER of \textit{case} 2 is close to BER of UL communication with no echo interference, because the echo signal interference endures large loss and is much smaller than the received UL communication signal. As $P_t^D$ increases, BER increases slowly before $P_t^D$ reaches 27 dBm. Because $P_t^D$ is not allowed to exceed 27 dBm in the 3GPP standard~\cite{3GPPV2X}, the echo signal in CDU JCAS causes negligible degradation of communication performance. 


\section{Conclusion}\label{sec:conclusion}
This paper proposes a CDU JCAS system that can improve the accuracy of environmental sensing for BS while performing reliable UL communication. A novel SIC-based CDU JCAS processing method is proposed to remove the interference of UL communication to echo sensing signal processing without reducing the reliability of UL communication. Extensive simulation results verify the feasibility of the CDU JCAS system. It is shown that the CDU JCAS processing method has significantly improved the range detection accuracy compared to the traditional DL active JCAS method, with negligible performance degradation of UL communications.

\begin{appendix}
\section{Solution to \eqref{equ:argmin_pro}} \label{Theo:A}
	It can be concluded that \eqref{equ:argmin_pro} is a convex problem of $\bf B$. To simplify the expression, we use ${\bf{h}}$, $P$, and $d$ to replace ${{\bf{h}}_{C,n,m}^U}$, $P_t^U$, and ${d_{n,m}^U}$ in \eqref{equ:argmin_pro}, respectively. By denoting
	\begin{equation}\label{equ:J}
		J = E\{ {\| {{{\bf{w}}^H}{\bf{h}}\sqrt P d - d} \|_2^2 + Tr( {{\sigma_N^2}{{\bf{w}}^H}{\bf{w}}} )} \},
	\end{equation} 
	the solution of $\frac{{\partial J}}{{\partial {\bf{B}}}} = {\bf{0}}$ is the minimum point for \eqref{equ:argmin_pro}. By deriving the first-order derivative of $J$ over $\bf B$, and exploiting $E\{ {\| {\bar d_{n,m}^U} \|_2^2} \} = 1$, we obtain
	\begin{equation}\label{equ:J_derivative}
		\frac{{\partial J}}{{\partial {\bf{B}}}} = 2{{\bf{h}}^H}{\bf{h}}P{{\bf{h}}^H}{\bf{hB}} - 2\sqrt P {{\bf{h}}^H}{\bf{h}} + 2{\sigma_N^2}{{\bf{h}}^H}{\bf{hB}}.
	\end{equation} 
	We can further derive the minimum point for \eqref{equ:argmin_pro} as 
	\begin{equation}\label{equ:B_solution}
		{\bf{B}} = {( {{{\bf{h}}^H}{\bf{h}}} )^{ - 1}}{( {P{{\bf{h}}^H}{\bf{h}} + {\sigma_N^2}} )^{ - 1}}( {{{\bf{h}}^H}{\bf{h}}} )\sqrt P.
	\end{equation} 
	By applying \eqref{equ:B_solution} into \eqref{equ:argmin_pro}, we obtain \eqref{equ:w_RX_C}.
\end{appendix}

{\small
	\bibliographystyle{IEEEtran}
	\bibliography{reference}
}
\vspace{-10 mm}

\ifCLASSOPTIONcaptionsoff
  \newpage
\fi

\end{document}